\documentclass[twocolumn,trackchanges]{aastex701}

\usepackage{amsmath}
\usepackage{float}
\usepackage{makecell}
\usepackage{threeparttable}
\usepackage{appendix}
\usepackage{booktabs}
\usepackage{xurl} 
\hypersetup{breaklinks=true}

\begin{document}

\title{Physical Origin of Very-High-Energy Gamma Rays from the Low-Luminosity Active Galactic Nucleus NGC 4278 and Implications for Neutrino Observations}

\author[]{Shilong Chen}
\affiliation{Laboratory for Relativistic Astrophysics, Department of Physics, Guangxi University,530004, Nanning , China }
\email[show]{slchen@ihep.ac.cn}  

\author[]{Abhishek Das} 
\affiliation{Department of Physics, The Pennsylvania State University, University Park, PA 16802, USA}
\affiliation{Department of Astronomy \& Astrophysics, The Pennsylvania State University, University Park, PA 16802, USA}
\affiliation{Center for Multimessenger Astrophysics, Institute for Gravitation and the Cosmos, The Pennsylvania State University, University Park, PA 16802, USA}
\email{abhidas5952@gmail.com}

\author[]{B. Theodore Zhang} 
\affiliation{Key Laboratory of Particle Astrophysics and Experimental Physics Division and Computing Center, Institute of High Energy Physics, Chinese Academy of Sciences, 100049 Beijing, China}
\affiliation{TIANFU Cosmic Ray Research Center, 610213,Chengdu, Sichuan , China}
\email[show]{zhangbing@ihep.ac.cn \\}

\author[]{Shigeo S. Kimura} 
\affiliation{Frontier Research Institute for Interdisciplinary Sciences, Tohoku University, Sendai 980-8578, Japan}
\affiliation{Astronomical Institute, Tohoku University, Sendai 980-8578, Japan}
\email[show]{shigeo@astr.tohoku.ac.jp}

\author[]{Kohta Murase} 
\affiliation{Department of Physics, The Pennsylvania State University, University Park, PA 16802, USA}
\affiliation{Department of Astronomy \& Astrophysics, The Pennsylvania State University, University Park, PA 16802, USA}
\affiliation{Institute for Gravitation and the Cosmos, The Pennsylvania State University, University Park, PA 16802, USA}
\affiliation{Center for Gravitational Physics and Quantum Information, Yukawa Institute for Theoretical Physics, Kyoto University, Kyoto 606-8502, Japan}
\email[show]{murase@psu.edu}

\author[]{Yunfeng Liang} 
\affiliation{Laboratory for Relativistic Astrophysics, Department of Physics, Guangxi University,530004, Nanning , China }
\email{liangyf@gxu.edu.cn}  

\begin{abstract}
Relativistic jets in active galactic nuclei (AGNs) are known to accelerate particles to extreme energies, yet the physical origin of very-high-energy (VHE) emission from low-luminosity AGNs (LL AGNs) remains unclear. NGC 4278, a  {nearby} LLAGN, has recently been identified as a VHE source following detections by LHAASO. In this  {work}, we present a multiwavelength and multi-messenger analysis to investigate the origin of this emission. Swift-XRT monitoring reveals a quasi-quiescent state characterized by  {the} low X-ray flux. Modeling the broadband spectral energy distribution with the leptohadronic code \textsc{AMES}, we find that a standard one-zone synchrotron self-Compton (SSC) model underpredicts the VHE flux, unless a  {relatively} high Doppler factor ($\delta \gtrsim  {6}$) is invoked. Alternatively, an external inverse-Compton (EIC) scenario—scattering seed photons from a radiatively inefficient accretion flow (RIAF)—provides a good description of the broadband  {emission} with modest jet power and Doppler factor. We further explore neutrino production within a leptohadronic framework.  {The EIC model in the quasi-quiescent state yields the largest predicted number of muon neutrinos, reaching $N_{\nu_{\mu}} \sim 0.001$ over 15 years of IceCube observations (assuming that 0.1\% of the Eddington luminosity is converted into high-energy protons).} Future multimessenger observations are essential to unveil the details of the high-energy processes of NGC 4278.
\end{abstract}

\keywords{\uat{High energy astrophysics}{739} --- \uat{Gamma-ray astronomy}{628}  --- \uat{Low luminosity active galactic nuclei}{2033} --- \uat{Galaxy jets}{601}--- \uat{Neutrino astronomy}{1100}}


\section{Introduction} 
Low-luminosity active galactic nuclei (LL AGNs) represent a major AGN population and are ubiquitous in the Universe~\citep{ho1999spectral,ho2008nuclear,2005A&A...435..521N}. 
They exhibit lower bolometric luminosities ($L_{\rm bol}\sim 10^{38}-10^{44}\rm~erg~s^{-1}$), relative to their Eddington luminosities~\citep{ho2008nuclear}.
 {In terms of their spectral properties, LL AGNs are characterized by the absence of a prominent big blue bump and Fe K$\alpha$ line, as well as the presence of flat or inverted radio spectra and weak optical/UV emission~\citep{ho1999spectral, ho2008nuclear}.}
The most plausible explanation for their low luminosity is a combination of low mass accretion rates and low radiative efficiencies, likely  {resulting from} an insufficient gas supply in the host galaxy. Consequently, these systems lack the bright broad-line region (BLR) and the optically thick accretion disk found in standard AGNs~\citep{ho2008nuclear}.

NGC 4278 is an elliptical galaxy belonging to the LGG 279 group
~\citep{1993A&AS..100...47G}. 
It is located at a redshift of $z = 0.00216$, with $\text{R.A.} = 185.03^\circ$ and $\text{Dec.} = 29.28^\circ$~\citep{gonzalez2009x}.
The galaxy hosts a supermassive black hole (SMBH) with a mass of $M_{\rm BH}=(3.09\pm 0.58)\times 10^8 M_{\odot}$ \citep{2003MNRAS.340..793W,2005ApJ...625..716C}, corresponding to an Eddington luminosity of $L_{\rm Edd}\simeq 3.9\times10^{46}\; \rm{erg \; s^{-1}}$. 
NGC 4278 is accreting at a low rate, with an Eddington ratio of $\lambda_{\rm Edd} \equiv L_{\rm bol}/L_{\rm Edd} \sim 7 \times 10^{-6}$ \citep{Younes_2010}.
Although NGC 4278 is also classified as a radio-loud AGN, its radio luminosity is at least two orders of magnitude lower than that of powerful radio-loud AGNs.  
Based on the measured radio core luminosity, the jet power is estimated to be $P_{\rm jet} \sim (1-2)\times10^{42}{\rm~{erg~s^{-1}}}\ll L_{\rm Edd}$ \citep{2005ApJ...622..178G,2010ApJ...720.1066C,pellegrini2012agn}. 

The first LHAASO catalog reported the detection of 1LHAASO J1219+2915, a new high-latitude TeV source identified with the nearby LL AGN NGC 4278~\citep{LHAASO:2023rpg}. Detected at a significance of 8.8$\sigma$, the source is spatially coincident with NGC 4278 within $0.03^\circ$, strongly supporting a physical association~\citep{LHAASO:2024qzv}. 
This detection is notable as NGC 4278 is a LINER galaxy  {with no gamma-ray counterpart in the Fermi-LAT 4FGL-DR4 catalog}~\citep{Abdollahi_2022,ballet2024fermilargeareatelescope}.
The origin of VHE gamma rays detected by LHAASO during the flaring state has been  {interpreted within} the one-zone synchrotron self-Compton (SSC) model \citep{lian2024originhighenergygammarays,Dutta_2024,2024ApJS..271...10W}. However, the origin of VHE gamma rays in the quasi-quiescent state remains unclear. 
Recently, \cite{Shoji:2025znc} proposed that these gamma rays originate from $pp$ interactions of escaped cosmic rays (CRs). 
In this study, we analyze publicly available X-ray and gamma-ray data from 2024-2025, as well as public neutrino data, and explore their implications for the origin of the VHE emission.

This paper is structured as follows: Section~\ref{sec:obs} presents a detailed analysis of the broadband SED of NGC 4278, including X-ray, gamma-ray, and neutrino data. Section~\ref{sec:method} describes our methods and the fitting procedure using a leptohadronic numerical code. The results are presented in Section~\ref{sec:result}, followed by a discussion and summary in Section~\ref{sec:sum}.

\section{Multiwavelength and multimessenger observations}\label{sec:obs}
\subsection{Radio observations}
VLBA and VLA observations of NGC 4278 at 5 GHz and 8.4 GHz in 1995 and 2000  {revealed} a two-sided jet structure with total flux densities ranging from 95 to 135 mJy~\citep{2005ApJ...622..178G}.
Assuming that the observed asymmetry arises from Doppler beaming rather than environmental inhomogeneities, jet kinematics was constrained using the arm-length ratio and apparent separation velocity $\beta_{\mathrm{sep}}$. These measurements yield a Doppler factor of $\delta \sim 2.6-2.8$, indicating the jets are mildly relativistic and viewed at a small angle to the line-of-sight.
The jet extends approximately 20 mas, corresponding to a physical scale of $1.4\,\text{pc}$. The major axis of N2, one of the largest components within the radio emission region, measures 11.17 mas, which translates to a physical size of $0.81\text{pc}$.~\citep{2005ApJ...622..178G}.

\subsection{X-ray Analysis}
We analyze the Swift-XRT observations of NGC 4278 ( {Table~\ref{tab:xray_fit_error} and table~\ref{tab:xrt}} in the Appendix) using \texttt{HEASOFT} (v6.35.1). 
Data reduction was performed with \texttt{xrtpipeline} (v0.13.7) and \texttt{CALDB} (v20240522).
Source events  {were extracted using \texttt{xselect}} from a 20-pixel radius circle centered on the position of NGC 4278, while background events  {were} taken from a 30--45 pixel annulus. 
We  {generated} the corresponding Ancillary Response Files (ARFs) using standard tools.
Spectra  {were} fitted in the 0.3--8 keV band using \texttt{XSPEC} (v12.15.0) with C-statistics. We  {applied} an absorbed power-law model, fixing the Galactic column density at $N_{\rm H} = 2.2 \times10^{20}\rm~cm^{-2}$~\citep{2013MNRAS.431..394W} and allowing the intrinsic absorption, photon index, and normalization to vary.
A joint fit to the dataset yielded a photon index of $\Gamma = 2.11$ and an unabsorbed flux of $1.43 \times 10^{-12}~\rm{erg ~cm^{-2}~s^{-1}}$ in the 0.3--10 keV band. 
The detailed fitting results are presented in Table~\ref{tab:xray_fit_error} in the Appendix.
Comparison with archival observations~\citep{Younes_2010,pellegrini2012agn} confirms that the source remained in a quasi-quiescent state during this period.

\subsection{GeV Gamma-ray Analysis}
Although NGC 4278 is generally faint in GeV gamma rays, a transient detection ($4.3\sigma$) has been reported during the 2021--2022 LHAASO active phase~\citep{bronzini2024fermilatdetectionlowluminosityradio}. Here, we analyze Fermi-LAT data covering MJD 60310--60796 (Jan 2024 -- May 2025) using \texttt{Fermitools} (v2.4.0) and \texttt{Fermipy} (v1.4.0)\citep{2017ICRC...35..824W}.
We selected Pass 8 events (100 MeV--300 GeV) within a $15^\circ$ ROI( {region of interest}) centered on the source. Standard cuts (\texttt{zenith angle $<90^\circ$}, \texttt{DATA\_QUAL>0}) and the \texttt{P8R3\_SOURCE\_V3} IRFs were used.
The model included Galactic and isotropic diffuse backgrounds alongside 4FGL-DR4 point sources. Parameters for sources within $5^\circ$ are free, while those for sources between $5^\circ$--$10^\circ$  {had only their normalizations free}.
The analysis yields $\rm TS=8.46$, falling below the detection threshold ($\rm TS \ge 25$).  {We then divided the data into energy bins and perform a bin-by-bin likelihood fit with the spectral index fixed at 2 to derive the flux for each bin. Consequently, for bins with low significance, we compute 90\% confidence level upper limits using the profile likelihood method.  {The corresponding light curves are presented in Figure \ref{fig:light_curve}, while the upper limits in different energy bins are shown in Figure \ref{fig:data}}}

\subsection{VHE Gamma-ray Observations}
Leveraging the high sensitivity of LHAASO, the LHAASO Collaboration recently reported the detection of the VHE gamma-ray source 1LHAASO J1219+2915, spatially associated with NGC 4278,  {based on data collected} between 2021 March 5 and 2023 October 31~\citep{LHAASO:2024qzv}.
Historically, NGC 4278 has been classified as a radio-loud AGN but  {has} lacked significant gamma-ray detection in Fermi-LAT catalogs, leading to earlier assumptions that such low-luminosity AGNs might be inefficient particle accelerators at high energies.

\begin{figure*}
    \centering
\includegraphics[width=\linewidth]{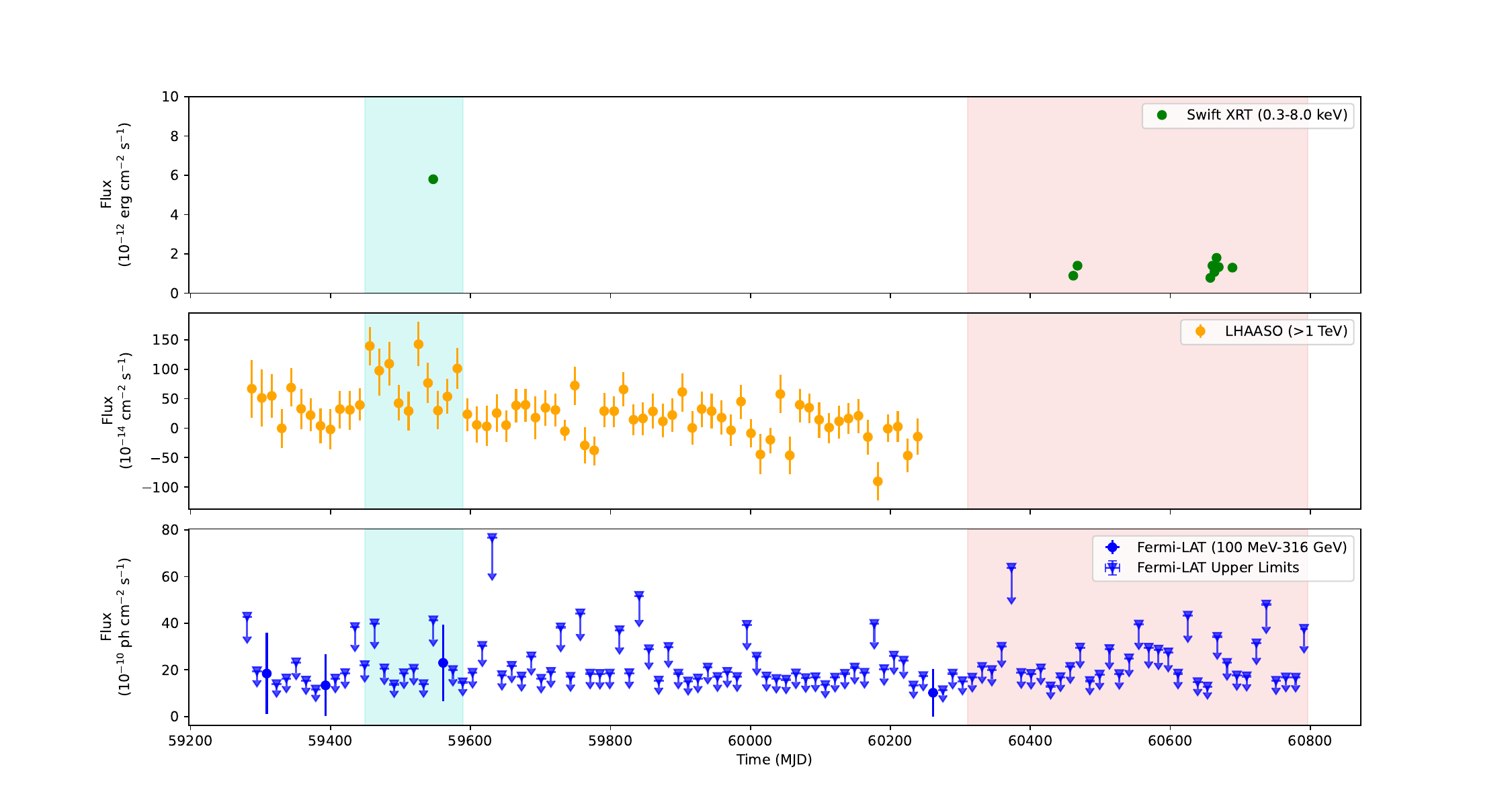}
    \caption{Multiwavelength light curves. We mainly  {focus} on two time periods. The first period (cyan band) corresponds to the LHAASO flaring state, from MJD 59449 to 59589 \citep{LHAASO:2024qzv}. The second period (pink band) spans from MJD 60310 to MJD 60796. For Swift-XRT data,  {we show the flux for each observation}. For the Fermi-LAT light curve, we use a time binning of 7 days.}
    \label{fig:light_curve}
\end{figure*}

The LHAASO analysis modeled the source spectrum using a power-law with attenuation from the Extragalactic Background Light (EBL).
The source exhibits distinct activity states: a flaring state with an integrated luminosity of $L_{0.1-10 \rm~TeV} \simeq 3.0\times10^{41}\rm~{erg~s^{-1}}$, and a quasi-quiescent state with $L_{0.1-10 \rm~TeV} \simeq 4.3\times10^{40}\rm~{erg~s^{-1}}$~\citep{LHAASO:2024qzv}. 
X-ray and GeV light curves covering the 2024--2025 epoch are presented in Figure~\ref{fig:light_curve}. The lack of significant variability in these bands supports the classification of the source as in a quasi-quiescent state during this period.

\subsection{Neutrino Observations}
\label{Neutrino observation}
For this work, we perform an unbinned likelihood analysis,  {following the methodology used in previous IceCube point-source searches} ~\citep{https://doi.org/10.21234/cpkq-k003,2018,2022,2008APh....29..299B} to evaluate the significance of a potential neutrino signal originating from the direction of NGC 4278. 
Both time-integrated and time-dependent methods are employed to search for neutrino emission from the source. 

\begin{table*}[htbp]
\caption{\label{tab:llagn} {Results of the analysis NGC 4278 neutrino flux by IceCube's 10 year public data.}} 
\begin{ruledtabular}
\begin{tabular}{lcccccc}
Method & $n_s$ & $\gamma$ & $\textrm{TS}$ & $-\textrm{log}10(p_{\rm local})$&$\Phi_{0}$ at $1\rm~TeV$  ($\rm{TeV^{-1}s^{-1}cm^{-2}}$)\\ \hline
Time integration  & 11 & 3.6 & 0.83 & 0.48 & $3.7 \times 10^{-13} (\rm{90\%~Sensitivity})$ \\
 Time dependent  & 2.0 & 1.9 & 8.2 & 0.69 & $8.0 \times 10^{-13} $(90\%~Upper Limit) \\
\end{tabular}
\end{ruledtabular}
\end{table*}
We utilize the software \texttt{SkyLLH\footnote{https://icecube.github.io/skyllh/master/html/index.html}},  {a Python-based tool for developing} such analyses in a telescope-independent framework \citep{Bellenghi:20230u,Abbasi:2021s7,Wolf:2019Pu}  {which we use to analyze IceCube’s 10-year public data}~\citep{2018}. We use the time-integration and time-dependent methods to search  {for} neutrino signal, respectively. The results are summarized in Table~\ref{tab:llagn}. In the time-integrated analysis, although the best-fit value of $n_s = 11.4$ is obtained, the corresponding Test Statistic (TS) is negligibly small. This low TS indicates that the apparent excess is consistent with background fluctuations; thus, no significant neutrino signal is detected. As a reference, the time-integrated 90$\%$ sensitivity of IceCube with a spectral index of 2.0 is approximately $3.7\times10^{-13}\,\mathrm{TeV^{-1}s^{-1}cm^{-2}}$ at 1 TeV \citep{IceCube:2019cia}. For the time dependent method,  {we obtain an excess of 2.0 neutrino events}, corresponding to the TS of 8.2. The best-fit time is $56704.10\pm1.18$ (MJD). However, it does not reach the level of a statistically significant detection.  {We perform trials to derive the upper limit on the neutrino flux by injecting different numbers of signal events. The neutrino spectrum is modeled with a simple power-law function, where $\Phi_0$ is the normalization at 1~TeV and $\Gamma$ is the spectral index:
\begin{equation}
\Phi(E)=\Phi_{0}\left(\frac{E}{1000~\mathrm{GeV}}\right)^{-\Gamma}.
\end{equation}
This leads to a neutrino flux upper limit of $\sim 8 \times 10^{-13}~ \rm{TeV^{-1}~cm^{-2}~s^{-1}}$ at 1~TeV.}

\begin{figure}[htbp]
\includegraphics[width=\linewidth]{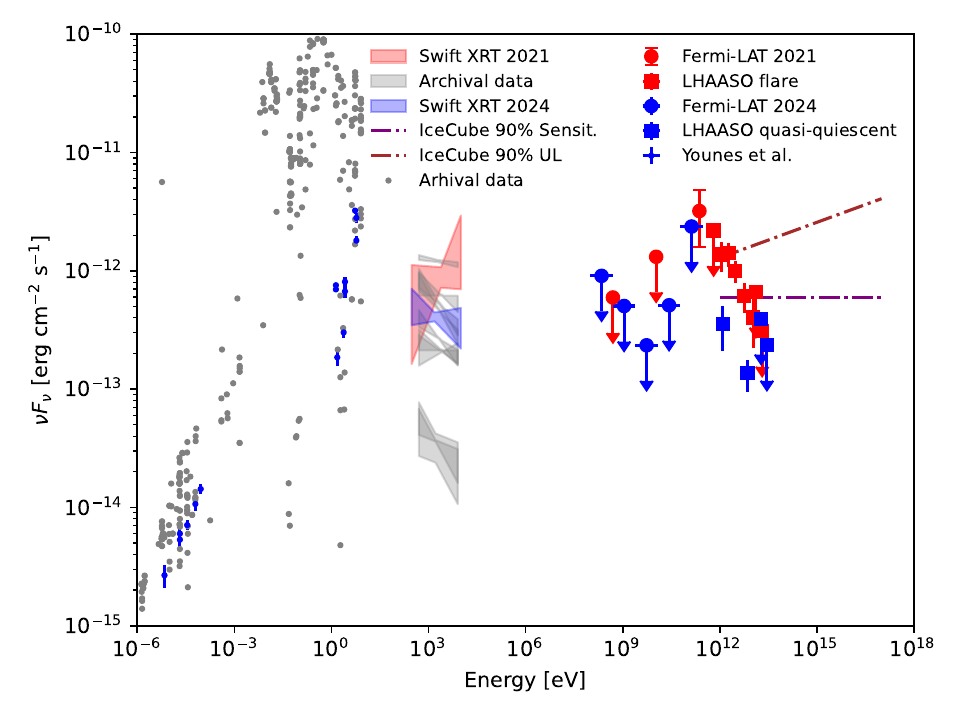}
\caption{Broadband SED of NGC 4278. The archival data (small blue points) \citep{Younes_2010} correspond to radio and optical observations.  {Additional archival data\citep{2003MNRAS.341....1M,2007ApJS..171...61H,1970ApJS...20....1D,1997ApJ...475..479W,2007MNRAS.376..371J,1998AJ....115.1693C,1990IRASF.C......0M,1994yCat.2125....0J,1996ApJS..103..427G,1992ApJS...79..331W,2010AJ....140.1868W} (small gray points) are obtained from} SSDC\footnote{https://tools.ssdc.asi.it/SED/}. The archival X-ray spectrums (gray band) are obtained from Chandra and XMM-Newton observations \citep{pellegrini2012agn}, where the upper band corresponds to the XMM-Newton observation in 2004 and the lower band to the Chandra observation in 2007. The X-ray spectrum (pink band) \citep{lian2024originhighenergygammarays} is from Swift–XRT observations in 2021, while the quasi-quiescent X-ray state (blue band) is from Swift–XRT observations in 2024. The GeV data points (red squares) \citep{bronzini2024fermilatdetectionlowluminosityradio} are from Fermi–LAT observations in 2021, and the blue squares represent our result which are obtained by analyzing Fermi–LAT data from 2024. The VHE data (red and blue dots) \citep{LHAASO:2024qzv} are obtained from LHAASO observations during an active phase of this source. The brown dash-dotted line is the neutrino 90$\%$ sensitivity of IceCube with a spectral index of 2.0\citep{IceCube:2019cia}. The purple dash-dotted line is the neutrino 90$\%$ upper limit from this work.}
\label{fig:data}
\end{figure}

In Figure~\ref{fig:data}, we summarize the broadband SED of NGC 4278 in both quasi-quiescent and flaring states, including the neutrino sensitivity. 

\section{Physical origin of multiwavelength emission}\label{sec:method}
\subsection{Particle transport and radiative processes}
We model the spectral evolution of the jet by solving time-dependent, coupled transport equations. In this work, we solve Equations~\ref{eq:fp_e} and \ref{eq:fp_p} using the AGN module of {\sc AMES}~\citep[e.g.,][]{Murase:2022dog,Zhang_2023,Das:2024vug}.

\begin{align}
\frac{\partial n_{\varepsilon_e}^{e}}{\partial t}
&= - \frac{n_{\varepsilon_e}^{e}}{t_{\mathrm{esc}}^{e}}
   - \frac{\partial}{\partial \varepsilon_e}
     \left[
       \left(
          P_{\mathrm{syn}}^{e}
         + P_{\mathrm{IC}}^{e}
       \right)
       n_{\varepsilon_e}^{e}
     \right] \nonumber \\
&\quad
   + \frac{\partial}{\partial t}
     \left(
       n_{\varepsilon_e}^{\gamma\gamma}
       + n_{\varepsilon_e}^{\mathrm{BH}}
       + n_{\varepsilon_e}^{\mathrm{phmes}}
       + n_{\varepsilon_e}^{\beta\text{-}\mathrm{dec}}
     \right)
   + \dot{n}_{\varepsilon_e}^{\mathrm{inj}} .
\label{eq:fp_e}
\end{align}

\begin{align}
\frac{\partial n_{\varepsilon_\gamma}^{\gamma}}{\partial t}
&=
- n_{\varepsilon_\gamma}^{\gamma}
\left(
\frac{1}{t_{\rm esc}^{\gamma}}
+ \frac{1}{t_{\gamma\gamma}}
\right)
\nonumber \\
&\quad
+ \frac{\partial}{\partial t}
\left(
n_{\varepsilon_\gamma}^{\rm syn}
+ n_{\varepsilon_\gamma}^{\rm IC}
+ n_{\varepsilon_\gamma}^{\rm phmes}
\right)
+ \dot{n}_{\varepsilon_\gamma}^{\rm inj} .
\label{eq:fp_p}
\end{align}

The time evolution of the electron energy distribution $n^{e}_{\varepsilon_e}$ is described by the transport equation that includes particle escape, energy losses, secondary particle injection, and primary electron injection. The first term on the right-hand side accounts for electron escape with a characteristic timescale $t^{e}_{\rm esc}$. The second term describes energy losses due to synchrotron radiation and IC scattering, with loss rates $P^{e}_{\rm syn}$, $P^{e}_{\rm IC}$, respectively. The third term represents the injection of secondary electrons and positron pairs produced via $\gamma\gamma$ pair production, Bethe-Heitler pair production, photomeson production interactions, and neutron $\beta$-decay. Finally, $\dot{n}^{\rm inj}_{\varepsilon_e}$ denotes the primary electron injection rate.
The time evolution of the photon number density $n^{\gamma}_{\varepsilon_{\gamma}}$ is governed by the transport equation. The loss terms include photon escape from the emission region and absorption via $\gamma\gamma$ interactions, while the source terms arise from synchrotron emission, IC scattering, photohadronic-induced cascades, and primary photon injection.

The acceleration process is parametrized by a characteristic timescale $t_{\rm acc} \approx \eta \frac{\varepsilon^\prime}{e B^\prime c}$,
where $\varepsilon^\prime$ is the particle energy in the comoving frame, $B^\prime$ is the magnetic field strength, and $\eta \ge 1$ is a factor characterizing the acceleration efficiency (with $\eta = 1$ corresponding to the Bohm limit)~\citep{2015SSRv..191..519S}.
For photons, the escape timescale is the light-crossing time of the emission region, $t_{\rm esc, \gamma} = R^\prime / c$. For charged particles,  {given that the system is relativistic, we assume the similar escape timescale, $t_{\rm esc} = R^\prime/c$.}

The escape of charged particles from the blob is often described by diffusive processes, with $t_{\rm esc} \approx {R^\prime}^2 / 2 D$, where $D$ is the diffusion coefficient. Here, we consider advective escape, which can be approximated as $t_{\rm esc} \approx R^\prime / c$. This type of advection escape is dominant when the magnetic field is significantly amplified at the blob by some mechanism, and such  {an} amplified field quickly decays with a dynamical timescale of $R^\prime/c$. 
The adiabatic cooling is neglected, as we assume a cylindrical geometry and  {a} non-expanding blob.

We assume the primary electrons and protons are injected with a power-law spectrum featuring an exponential cutoff:
\begin{equation}
\dot{n}_{\varepsilon_e}^{\mathrm{inj}} \propto \varepsilon_e^{\prime -s} \exp\left(-\frac{\varepsilon_e^\prime}{\varepsilon_e^{\prime_{\rm max}}}\right),
\end{equation}
where $s$ is the spectral index and $\varepsilon_e^{\prime_{\rm max}}$ is the maximum cutoff energy.

For leptonic radiative processes, we consider synchrotron emission, IC scattering, and two-photon pair production.
The synchrotron cooling timescale is given by:
\begin{equation}
    t_{\rm syn} = \frac{p}{|\dot{p}|_{\rm syn}} \approx \frac{3 m^4 c^3}{4 \sigma_{\rm T} m_e^2 c^2} \frac{1}{\gamma \beta^2 U_B^\prime},
\end{equation}
where $U_B^\prime = {B^\prime}^2 / 8\pi$ is the magnetic energy density, $\sigma_{\rm T}$ is the Thomson cross-section.

The IC cooling timescale for relativistic electrons is expressed as:
\begin{equation}
    t_{\rm IC} \approx \frac{3 m_e c}{4 \sigma_{\rm T} U_{\rm ph}^\prime \gamma F_{\rm KN}},
\end{equation}
where $U_{\rm ph}^\prime$ is the target photon field energy density and $F_{\rm KN}$ is the Klein-Nishina correction factor~\citep{Jones:1968zza}. In our numerical calculations, we adopt the established results for the total and differential cross sections (e.g. \cite{1986rpa..book.....R}).

The optical depth for two-photon annihilation ($\gamma\gamma \to e^+e^-$) is calculated as:
\begin{equation}
    t_{\gamma\gamma}(\varepsilon_\gamma^\prime) = \frac{R^\prime}{c} \int d\Omega \int d\varepsilon_* \, n_{\rm ph}(\varepsilon_*, \Omega) \, \sigma_{\gamma\gamma}(\varepsilon_\gamma^\prime, \varepsilon_*, \mu) \, (1 - \mu) c,
\end{equation}
where $\varepsilon_*$ is the target photon energy, $\mu = \cos\theta$ is the interaction angle, and $\sigma_{\gamma\gamma}$ is the pair-production cross-section.

\subsection{Fitting Method}
We employ the MCMC software \texttt{emcee} (v3.1.6)~\citep{Foreman_Mackey_2013} to fit the broadband SED of NGC 4278  {with} {\sc AMES}. The best-fit parameters are listed in Table~\ref{tab:combined_parameters}.

The log-likelihood function ($\ln \mathcal{L}$) incorporates both detected data points and upper limits (ULs). For the X-ray, GeV, and VHE $\gamma$-ray data, we adopt the reported measurement uncertainties. 
 {In contrast, for the radio, infrared (IR), and optical datasets—which often have very small statistical uncertainties—we need to account for additional systematic uncertainties arising from absolute flux calibration. These uncertainties are added in quadrature to the statistical errors to ensure a balanced weighting across different energy bands in the SED fitting. The representative calibration uncertainties are $\sim$2\% for optical data~\citep{Bohlin:2014nwa}, $\sim$5\% for VLA radio observations~\citep{2013ApJS..204...19P}, and $\sim$5–10\% for VLBI measurements~\citep{Kovalev_2005}, consistent with typical instrument performance. However, the error in each spectral point should also reflect the flux variations during the observation period used~\citep{Hervet:2023rpe, Podlesnyi:2025aqb}. The radio flux shows modest variability on yearly time-scales with decreasing flux by about 10\% between 1995 to 2000~\citep{2005ApJ...625..716C, Younes_2010}. The variability of UV and optical emission can be larger than radio flux, and the flux increased by a factor of 2 on month timescales~\citep{2009A&A...508..641C}. 
As a conservative assumption, we added a 10\% fractional uncertainty as an overall systematic error\footnote{ {We have also verified that the fitting results remain consistent when only the calibration uncertainties of the radio, IR, and optical datasets mentioned in the main text are considered.}}.}
For upper limits, we treat them as left-censored data and compute the probability that the true flux lies below the reported limit using the Gaussian cumulative distribution function (CDF), $\Phi(z)$. A conservative systematic uncertainty of 10\% ($\sigma_i = 0.1y_{\rm UL}$) is assigned to the upper limits.
The total log-likelihood is defined as:
\begin{equation}
\ln \mathcal{L}_{\rm total} = \sum_{i} 
\begin{cases} 
-\frac{1}{2} \left[ \frac{(y_i - \mu_i)^2}{\sigma_i^2} + \ln(2 \pi \sigma_i^2) \right] & \text{for detections} \\
\ln \Phi\left( \frac{y_i - \mu_i}{\sigma_i} \right) & \text{for UL}
\end{cases},
\label{eq:loglikelihood_combined}
\end{equation}
where $y_i$ is the observed flux (or UL value), $\mu_i$ is the model-predicted flux, and $\sigma_i$ is the flux error (or assigned UL uncertainty).

\section{Results}\label{sec:result}
\subsection{One-zone SSC model}
\subsubsection{Quasi-quiescent  state}
\begin{figure}[]
\includegraphics[width=\linewidth]{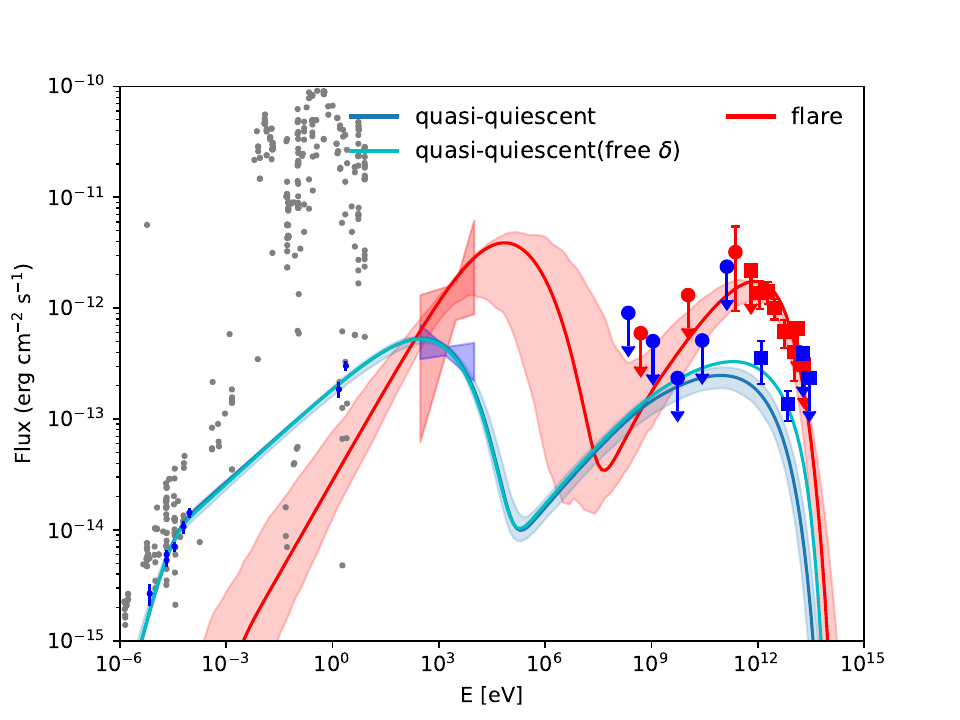}
\caption{The broadband SED modeled with one-zone SSC model. The solid red and blue lines represent the SED computed using the 50th percentile (median) of the parameter distributions for the flaring and quasi-quiescent  states, respectively. The surrounding red and blue bands indicate the 1 $\sigma$ uncertainty of the fitting parameters. 
}
\label{fig:spectrum-all}
\end{figure}

The SED in the quasi-quiescent state is well-reproduced by a one-zone leptonic SSC model, as shown in Figure~\ref{fig:spectrum-all}.
The shaded band represents the 1$\sigma$ uncertainty region derived from the MCMC fitting.
The contour plot is shown in the left panel  {of} Figure~\ref{fig:contour-fixed}.
The injected electron energy distribution is characterized by a spectral index of \( s_e \simeq 2.4 \), with minimum and maximum Lorentz factors of \( \gamma_{\min}^\prime \simeq 1.0\times10^3 \) and \( \gamma_{\max}^\prime \simeq 9.9\times10^{6} \), respectively. 
The magnetic field strength in the comoving frame is \( B^\prime \simeq 0.4~\mathrm{mG} \). 
The derived electron spectrum is noticeably softer than that inferred for the flaring state discussed later (see also~\cite{lian2024originhighenergygammarays,Dutta_2024,2024ApJS..271...10W}). 
We estimate the observed synchrotron peak energy as follows:
\begin{equation}
\varepsilon_{\mathrm{syn,pk}} \approx (1+z) \delta\frac{\gamma_e^2  \hbar e B'}{m_e c} 
    \simeq 1.2 \times 10^3\left(\frac{B'}{0.4\,\mathrm{mG}}\right) \left(\frac{\gamma_e}{10^7}\right)^2 \,\mathrm{eV},
\end{equation}
consistent with the modeled synchrotron peak shown in Figure~\ref{fig:spectrum-all}.

The corresponding total jet power in the black hole rest frame is estimated as \citep{Murase_2012}.
\begin{equation}
    P_{\rm jet} = 2\pi {R^\prime}^2 \beta c \Gamma^2 (u_B^\prime + u_e^\prime) \simeq  {6.5\times10^{42}~\mathrm{erg~s^{-1}}},
\end{equation}
which is approximately four times higher than the jet power inferred from radio observations~\citep{2005ApJ...622..178G}.
 {The size of the emission region is $R^\prime \sim 7.3\times 10^{17}\rm~cm$, which is larger than, but roughly consistent with, the size of the largest components within the radio emission region N2 mentioned above.}

\begin{table}[htbp]
\centering
\caption{Physical parameters derived from SSC modeling. Median values and $1\sigma$ confidence intervals for the SED model parameters of NGC 4278 during quasi-quiescent and flaring states.}
\begin{tabular}{lccc}
\hline
\hline
Parameter & \multicolumn{2}{c}{Quasi-quiescent} & Flare (2021) \\
~ & \multicolumn{2}{c}{2024-2025} & 2021 \\
\hline

 {$\delta$} &  {$2.7$[fixed]} &  {$6.3_{-1.58}^{+1.19}$} &  {$2.7$[fixed]} \\
$B' \,[\mathrm{mG}]$ & $0.40_{-0.10}^{+0.40}$ &  {$0.19_{-0.06}^{+0.19}$} & $29.80_{-26.30}^{+137.80}$ \\

$R' \,[10^{15}~\mathrm{cm}]$ & $731_{-486}^{+246}$ &  {$355_{-185}^{+276}$} & $2.03_{-1.68}^{+2.09}$ \\

$s_e$ & $2.43_{-0.03}^{+0.03}$ &  {$2.42_{-0.03}^{+0.03}$} & $1.88_{-0.24}^{+0.26}$ \\

$\gamma'_{e,\min}[10^{3}]$ & $1.01_{-0.72}^{+0.28}$ &  {$1.05_{-0.36}^{+0.37}$} & $1.05_{-0.83}^{+2.67}$ \\

$\gamma'_{e,\max}[10^{6}]$ & $9.94_{-3.11}^{+2.87}$ &  {$8.71_{-2.71}^{+3.04}$} & $18.7_{-10.7}^{+28.5}$ \\

$L'_e \,[10^{41}~\mathrm{erg\,s^{-1}}]$ & $39.2_{-19.0}^{+11.9}$ &  {$15.1_{-6.0}^{+10.6}$} & $1.45_{-0.70}^{+4.92}$ \\

\hline
\end{tabular}
\label{tab:combined_parameters}
\end{table}

As mentioned in the previous section, we initially fixed the Doppler factor at $\delta = 2.7$ when fitting the multiwavelength data. 
However, under this assumption, the predicted VHE $\gamma$-ray flux in the quasi-quiescent state is lower than the LHAASO measurements by a factor of $\sim 2$. 
We therefore  {treat} $\delta$ as a free parameter, and the resulting best-fit spectrum, shown in Figure~\ref{fig:spectrum-all}, provides a satisfactory description of the LHAASO observations, see the contour plots Figure~\ref{fig:contour-free} in the Appendix~\ref{coutour-fig}. 
 {For the best-fit model, the total jet power is estimated to be $P_{\rm jet} \simeq 7.3\times10^{43}~\mathrm{erg~s^{-1}}$ with $\delta \simeq 6.3$ and the best-fit radius of $R^\prime \sim 3.6 \times 10^{17}\,\text{cm}$.
Furthermore, Figure~\ref{fig:contour-free} in Appendix presents the best-fit spectrum obtained by varying $\delta$ in the range of 1--12 using the marginalized maximum-likelihood method. We  {also} find a preferred value of $\delta \sim 6$, consistent with the MCMC fitting results presented in Table~\ref{tab:combined_parameters}.
This suggests that a relatively high Doppler factor is required to reproduce the quasi-quiescent SED within the SSC scenario.}

We also evaluate the VHE gamma-ray origin within the context of external inverse-Compton (EIC) and leptohadronic scenarios in the following sections.

\subsubsection{Flaring State}
The SED fitting for the flaring state, using a one-zone leptonic SSC model, is represented by the red line in Figure~\ref{fig:spectrum-all}. This fit incorporates X-ray, $\gamma$-ray, and LHAASO data. Following the approach used for the quasi-quiescent  state, the Doppler factor  {is} fixed at $\delta = 2.7$. The fitted blob radius is $R^\prime \simeq 2.0 \times 10^{15}\rm~cm$. This comoving size can be independently estimated from the observed variability as $R^\prime \approx \delta c t_{\rm var}/(1+z) \simeq 4.1 \times 10^{17}\rm~cm$, adopting $t_{\rm var} = 58$ days \citep{LHAASO:2024qzv}.  {Our fitted value for $R^\prime$ is significantly smaller than this variability-derived upper limit, implying a more compact emission region.} The magnetic field strength  {is} found to be $B^\prime \simeq 30\rm~mG$. For the electron distribution, we  {obtain} a spectral index $s_e \simeq 1.9$, Lorentz factors $\gamma_{\rm min}^\prime \simeq 1.1 \times 10^3$ and $\gamma_{\rm max}^\prime \simeq 1.9 \times 10^7$, and a luminosity $L_e^\prime \simeq 1.5 \times 10^{41}\rm~erg~s^{-1}$. These parameters are summarized in Table~\ref{tab:combined_parameters}.
The total power of the two-sided jet is $P_{\rm jet} \simeq 2.5 \times 10^{41}\rm~erg~s^{-1}$, which is lower than the jet kinetic power derived from radio measurements \citep{2005ApJ...622..178G}. We estimate the synchrotron peak energy as: $E_{\rm syn,pk} \simeq 3.1 \times 10^5 (B^\prime/29.8 {\rm~mG}) (\gamma_e^2/10^7)\rm~eV.$ 
However, due to the lack of hard X-ray data, the synchrotron peak energy remains difficult to constrain precisely.
Our fitting results are generally consistent with those of \cite{lian2024originhighenergygammarays} and \cite{Dutta_2024}, though our derived emission region radius is smaller by a factor of approximately five. In their study, \cite{lian2024originhighenergygammarays} modeled the SED using a broken power-law electron distribution and  {an assumed radius, $R^\prime = 1 \times 10^{16}\rm~cm$}. While they also fixed $\delta = 2.7$ based on VLBA observations \citep{2005ApJ...622..178G}, they explored an alternative scenario with $\delta = 10$ based on TeV BL Lac studies \citep{2010MNRAS.401.1570T, Zhang_2012}. The discrepancies between our results and those of \cite{lian2024originhighenergygammarays} likely stem from the use of different Fermi-LAT datasets; specifically, we adopt the results from \cite{bronzini2024fermilatdetectionlowluminosityradio}, which provide more stringent {upper limits}.

\subsection{EIC Model}
In this section, we evaluate the contribution of external photon fields—specifically from the accretion disk—to the production of high-energy emission. While the SSC model considers internal synchrotron photons, the upscattering of external photons in the EIC model can also significantly contribute to the VHE gamma-ray flux. 
It has been widely discussed that the observed X-ray emission from  {LL AGNs originates from} the hot accretion  {flow}. Radiatively inefficient accretion flows (RIAFs) in LL AGN could produce multiwavelength emission  {extending into the X-ray band} by thermal electrons~\citep{Das:2026aed}, and we utilize the one-zone model provided by \citep{Kimura:2019yjo,Kimura_2020,Kimura_2021,Kimura_2021_b}.
In the comoving frame of the blob, the target photon density from the inner disk is de-boosted by the relativistic motion of the jet. The comoving photon energy density can be estimated as: 
\begin{equation}
{\varepsilon^\prime}^2\frac{dn}{d\varepsilon^\prime} = \delta_{D,\rm de}^4 \frac{d_L^2 EF_E }{l^2 c} ,
\end{equation}
where $l$ represents the distance from the SMBH to the emission blob measured in SMBH  {rest} frame, $\delta_{D,\rm de} =1/\Gamma(1+\beta) $, $\varepsilon^{\prime} = \delta_{D, \rm de} E (1+z)$.
We estimate the distance  {as} $l \approx R^\prime / \delta_D  \theta_j$, where $\delta_D = 1 / \Gamma(1-\beta) $ is the Doppler factor.
The jet velocity is $\beta = 0.76$, and the half-opening angle is assumed to be $\theta_j = 20^\circ$.
  The intrinsic jet opening angle is expected to be inversely proportional to the Lorentz factor, as predicted by both hydrodynamical and magnetic acceleration models~\citep{1979ApJ...232...34B, Komissarov:2007xx}. 
  In addition, the causal connection condition for relativistic jets requires $\Gamma \theta_j \lesssim 1$ for jets with equipartition fast magnetosonic speed and $\Gamma \theta_j \lesssim 0.7$ for jets with relativistic sound speed~\citep{2008ApJ...679..990Z, Clausen-Brown:2013mda}. In our case, $\Gamma \theta_j \approx 0.54$, ensuring that the jet remains causally connected. The empirical relation inferred from observations, $\Gamma \theta_j \sim 0.18$~\citep{Pushkarev:2017fbk}, also indicates that AGN jets are causally connected.
We have checked the EIC scenario for varying opening angles and found that, once the half-opening angle satisfies $\theta_j \gtrsim 7^\circ$, our results are not significantly affected. Although the photon field from the accretion disk is inherently anisotropic, we assume an isotropic distribution in our calculations, under the assumption that the high-energy electrons are isotropically distributed in the comoving frame. 
The discrepancy introduced by neglecting the full anisotropic IC scattering cross section is expected to be approximately a factor of two \citep{Zhang_2023}.

\begin{figure}[htbp]
\includegraphics[width=\linewidth]{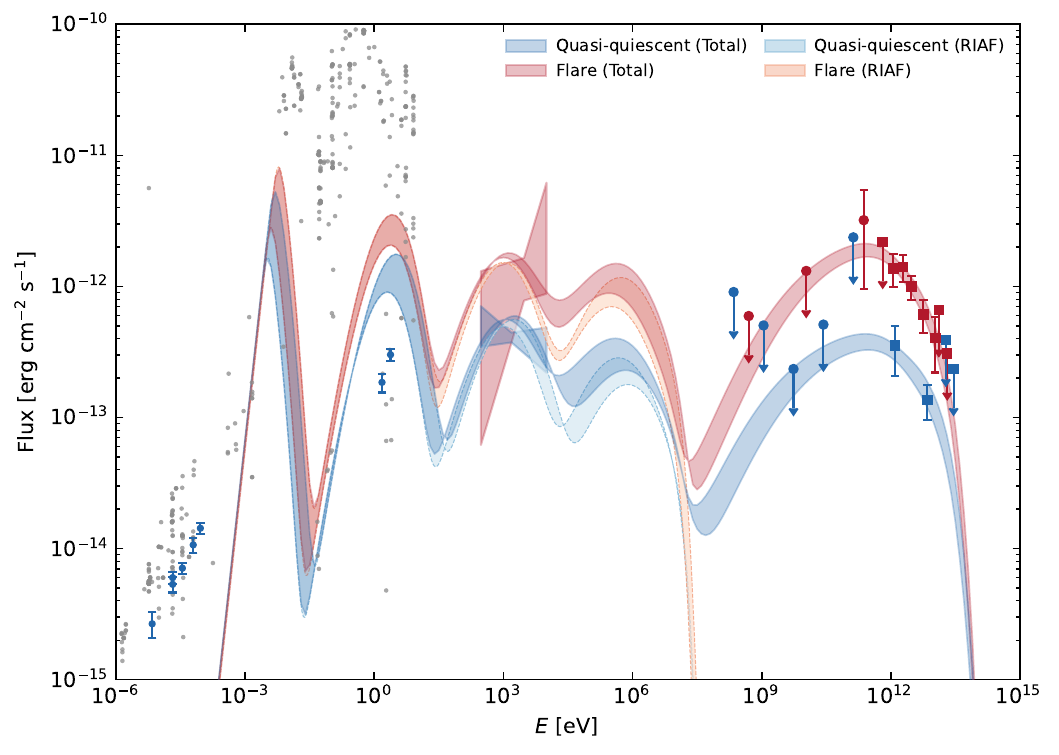}
\caption{ {Multiwavelength emission predicted by the EIC model. The dark band represent the total emission, while the light band denote the contributions from the RIAF.}
}
\label{fig:EIC_fit_sum_extra}
\end{figure}

Our results are shown in Figure~\ref{fig:EIC_fit_sum_extra}.
The dashed curves are the SED from our RIAF model~\citep{Kimura:2019yjo,Kimura_2020,Kimura_2021,Kimura_2021_b}, in both flaring and quasi-quiescent states, where the difference is mainly due to accretion rate.  {We adopt a set of best-fit EIC parameters corresponding to the lower boundary of the model uncertainty band shown in Figure~\ref{fig:EIC_fit_sum_extra}. 
The accretion rate is $\dot{m} = 1.1 \times 10^{-3}$ in the quasi-quiescent state, and it is $\dot{m} = 1.5\times10^{-3}$ in the flaring state. The other parameters are kept the same, with the alpha parameter $\alpha = 0.4$ and plasma beta parameter $\beta = 1.7$.} {We also use a set of EIC-preferred parameters, corresponding to the upper boundary of the model uncertainty band shown in  Figure~\ref{fig:EIC_fit_sum_extra}. The accretion rate is $\dot{m} = 9.0 \times 10^{-4}$ in the quasi-quiescent state, and it is $\dot{m} = 1.8\times10^{-3}$ in the flaring state. The alpha parameter is $\alpha = 0.6$ and plasma beta parameter $\beta = 0.3$.}
The best-fit parameters are summarized in Table~\ref{tab:EIC}.
In the EIC scenario, a harder spectral index of  electrons is injected to explain VHE gamma rays, while the X-ray emission is fully explained by the RIAF model.
 {Compared} to the SSC model, the EIC model provides a more reasonable explanation of the production of VHE gamma rays in the quasi-quiescent state for the fixed value of Doppler factor $\delta = 2.7$.
The corresponding total jet power is estimated as  {
$P_{\rm jet} \simeq 2.1\times 10 ^{39}~\mathrm{erg~s^{-1}}$ (quasi-quiescent ) and $P_{\rm jet} \simeq 7.0\times 10 ^{39}~\mathrm{erg~s^{-1}}$ (flare)},  {both of which} are smaller than the total jet power inferred from radio observations.
 {
We estimate the blob radius to be $R^\prime \sim 6.4 \times 10^{15}\,\text{cm}$ and $1.1 \times 10^{16}\,\text{cm}$ for the quasi-quiescent and flaring states, respectively. The size of the blob in the flaring state is approximately a factor of two larger than in the quasi-quiescent state; however, we consider this change to be modest and within the range of parameter uncertainties.} 
Both values are compact compared to the radio emission region of the jet.
The derived size of the emission region in the flaring state is a slightly larger than that of the quasi-quiescent state. While flares are conventionally attributed to highly compact regions, our EIC model fitting indicates a scenario wherein the flare occurs at radius close to the dominant quasi-quiescent emission zone.
{There are no direct observational constraints on the jet opening angle of NGC 4278 \citep[e.g.,][]{2005ApJ...622..178G}. However, a wide jet opening angle is not uncommon. For example, in the LLAGN M87, the jet opening angle is observed to be as large as $\sim 30^\circ$--$60^\circ$ at distances of $\sim 0.04$--$0.1$ pc from the central engine, representing the broadest opening angle reported for any extragalactic jet \citep{1999Natur.401..891J, Hada:2013yla}. Since the SMBH in NGC 4278 is approximately 20 times less massive than that in M87, a similar jet opening angle may be expected at distances of $\sim 0.002$--$0.05$ pc from the central engine. This range is broadly consistent with the location of the emission region inferred in the EIC model.}

 {The simulated spectrum exhibits a low-energy ``bumpy'' feature that is not typically observed in standard AGN SEDs or our archival data. This non-smooth structure likely arises from the standard formalism in our RIAF model, and similar features have been noted in previous studies of Sgr A* \citep{2003ApJ...598..301Y} and low-luminosity AGNs \citep{Nemmen:2013mya}. While distinct electron populations naturally produce discrete radiative components, a more rigorous multi-dimensional radiative transfer treatment—specifically including inverse Compton scattering by lower-temperature electrons—would bridge these gaps and yield a smoother SED.}
  {
We explicitly examine the smoothed spectrum and confirmed that our results are not affected by the more accurate Comptonization Kernel. 
The ``bumpy'' feature is expected to persist to some extent even with a more accurate Comptonization kernel, although its amplitude would be reduced~\citep{Quataert:1998yn}. 
A more careful treatment of the Comptonization kernel for the AGN SED and its effect on the predicted EIC emission will be addressed in future work.
}


\begin{table}[htbp]
\centering
\caption{ {Similar to Table~\ref{tab:combined_parameters}, physical parameters used in EIC modeling.}}
\begin{tabular}{lcc}
\hline
Parameter & quasi-quiescent & flare \\
\hline
$\dot{m}$ [fixed] &  {$1.1\times 10^{-3}$}&  {$1.5\times 10^{-3}$} \\
$\Gamma$ [fixed] & 1.5 & 1.5 \\
$\theta$ [deg, fixed] & 3 & 3 \\
$\theta_j$ [deg, fixed] & 20 & 20 \\
$B' \,[\mathrm{mG}]$ & 42 & 27 \\
$R^{\prime}\,[\mathrm{cm}]$ & $6.43\times10^{15}$ & $1.06\times10^{16}$ \\
$s_e$ & 1.60 & 1.53 \\
$\gamma^{\prime}_{e,\min}$ & $6.55\times10^{2}$ & $2.83\times10^{2}$ \\
$\gamma^{\prime}_{e,\max}$ & $2.87\times10^{7}$ & $2.00\times10^{7}$ \\
$L^{\prime}_{e}\,[\mathrm{erg\,s^{-1}}]$ & $2.80\times10^{39}$ & $1.22\times10^{40}$ \\
\hline
\end{tabular}
\label{tab:EIC}
\end{table}


\subsection{Leptohadronic model}
In addition to pure leptonic processes, VHE gamma rays can  {also be} produced  {by} protons, including proton synchrotron emission, decay of neutral pions, and associated cascade emission from the $pp$ and $p\gamma$ processes~\citep[e.g.,][]{Dermer:2012rg,2012A&A...546A.120D,2013ApJ...768...54B, Gao_2017, Keivani:2018rnh,Zhang:2019htg,Murase:2022dog}.
Nevertheless, in this study, we adopt the best-fit leptonic results as the baseline parameters for our leptohadronic calculations, rather than scanning the entire parameter space for leptohadronic modeling as in the previous section. This means that  {the} entire photon spectrum is described by the leptonic process, but we introduce  {a certain amount of protons} to evaluate the neutrino production in this system.

Similar to the electrons, the injected protons are assumed to follow a power-law energy distribution with an exponential cutoff, characterized by a fixed spectral index of $s_p = 2$. 
The maximum proton energy is determined by balancing the acceleration timescale with the diffusive timescale in the Bohm limit, yielding a maximum Lorentz factor of $\gamma_{p,\rm max}^\prime m_p c^2 \sim \eta^{-1} eB^\prime R^\prime$, where $\eta = 10$ is a numerical factor. 
 {The estimated values of $\gamma_{p,\rm max}^\prime$ are $9.9\times 10^6$ (SSC, quasi-quiescent), $1.9\times10^{6}$ (SSC, flare), $8.6\times10^{6}$ (EIC, quasi-quiescent) and $9.3\times10^{6}$ (EIC, flare).}
The minimum proton energy is $\varepsilon^\prime_{p, \rm min} = 10^9\rm~eV$.
The injected proton luminosity is fixed at $L_p^\prime = 3.9 \times 10^{43}\,\text{erg s}^{-1}$, roughly $0.1\%$ of the Eddington luminosity, which is consistent with the energy budget estimated by RIAF modeling in the EIC scenario.  {It is important to emphasize that this hadronic component is physically constrained by the system's available energy budget rather than the gamma-ray observational limits. Due to this stringent energetic restriction, the predicted hadronic cascade emission is negligible and naturally lies well below the observed broadband SED.}
The observed jet power of NGC 4278 is estimated to be only $\sim 10^{42}\,\text{erg s}^{-1}$, but we adopted the value above based on our RIAF modeling. Our choice of jet power is rather conservative,  {compared} the neutrino emission modeling in blazars \citep[e.g.,][]{Murase:2014foa,Dermer:2014vaa, Petropoulou:2015upa}, where the jet power close to the Eddington luminosity is often adopted.

\begin{figure*}
    \centering
    \includegraphics[width=0.4\linewidth]{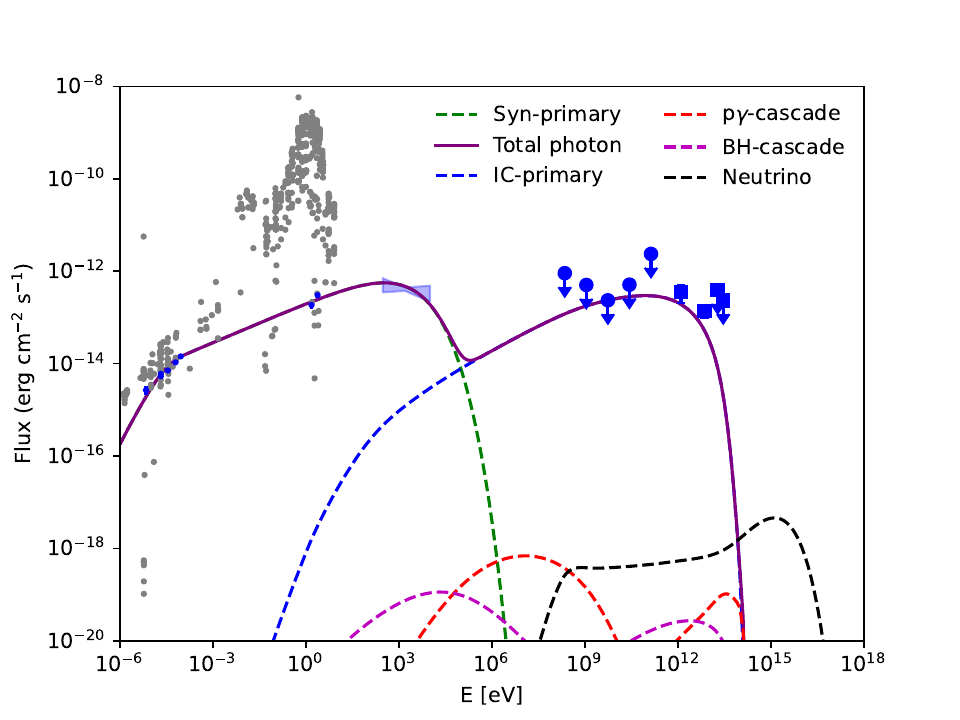}
    \includegraphics[width=0.4\linewidth]{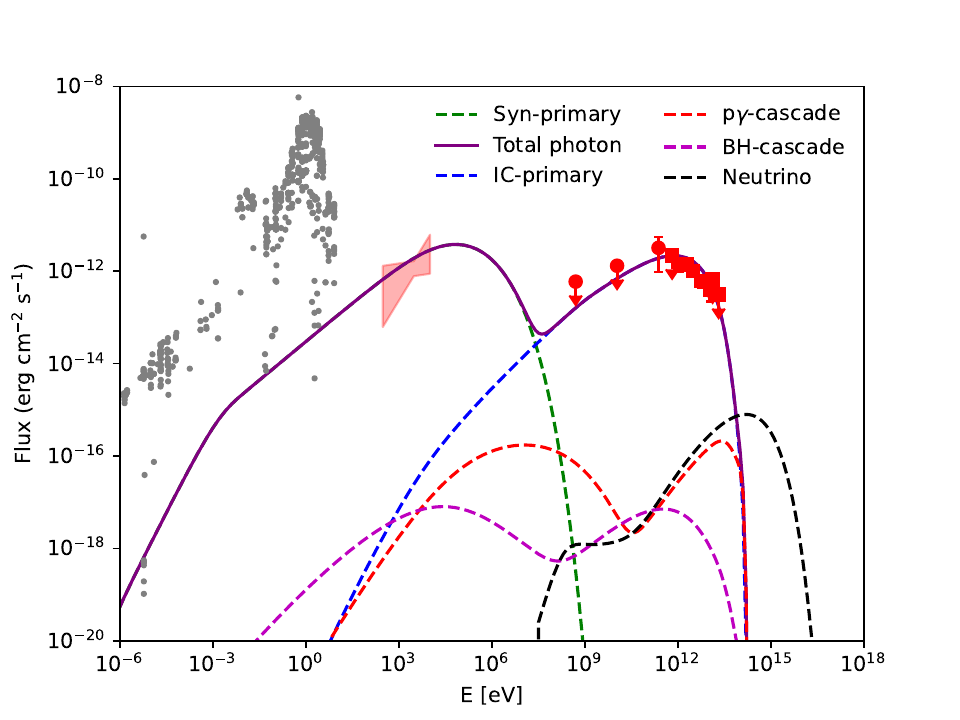}
    \includegraphics[width=0.4\linewidth]{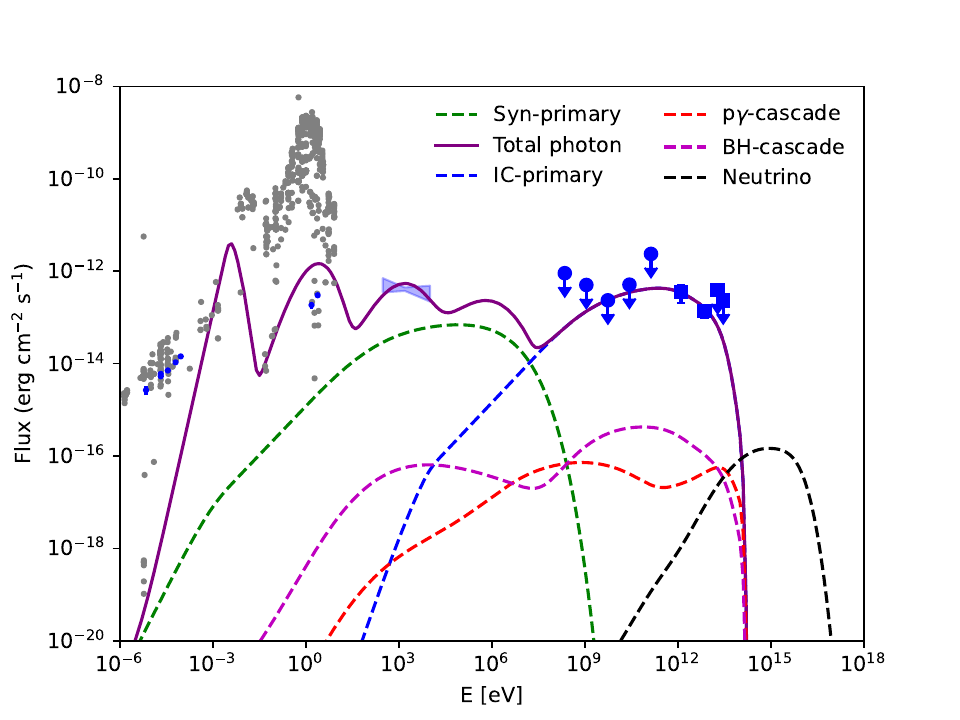} 
    \includegraphics[width=0.4\linewidth]{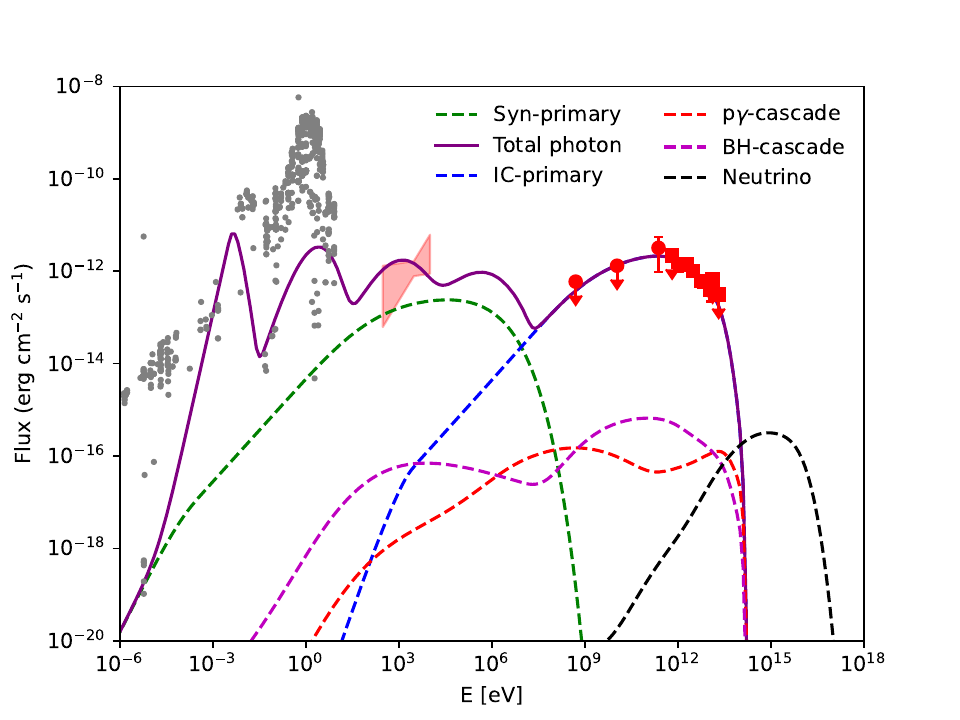}
\caption{ {Leptohadronic model with the additional injection of high-energy protons. The black dashed lines show the all-flavor neutrino spectrum, while the green, blue, red, and purple dashed lines represent the synchrotron emission from primary electrons, IC emission from primary electrons, cascade emission from the photomeson production process, and cascade emission from the Bethe-Heitler pair production process, respectively. Top left: quasi-quiescent state (SSC); Top right: flaring state (SSC); Bottom left: quasi-quiescent state (EIC); Bottom right: flaring state (EIC).}} 
    \label{fig:leptohadronic-cascade}
\end{figure*}

We include photomeson production, photopair production, and proton-proton ($pp$) interaction processes in the calculation. 
The number density of cold protons in the comoving frame is assumed to be equal to that of the electrons, 
$n_{p,\rm cold}^\prime = n_e^\prime \simeq L_e^\prime / (4\pi {R^\prime}^2 c \gamma_{e,\rm min}^\prime m_e c^2)$, and these protons serve as the targets for $pp$ interactions.
The detailed SEDs are presented in Figure~\ref{fig:leptohadronic-cascade}. The cascade emissions from photomeson production and Bethe-Heitler pair production are shown separately; both lie well below the observed SED.

\begin{figure}
    \centering
    \includegraphics[width=\linewidth]{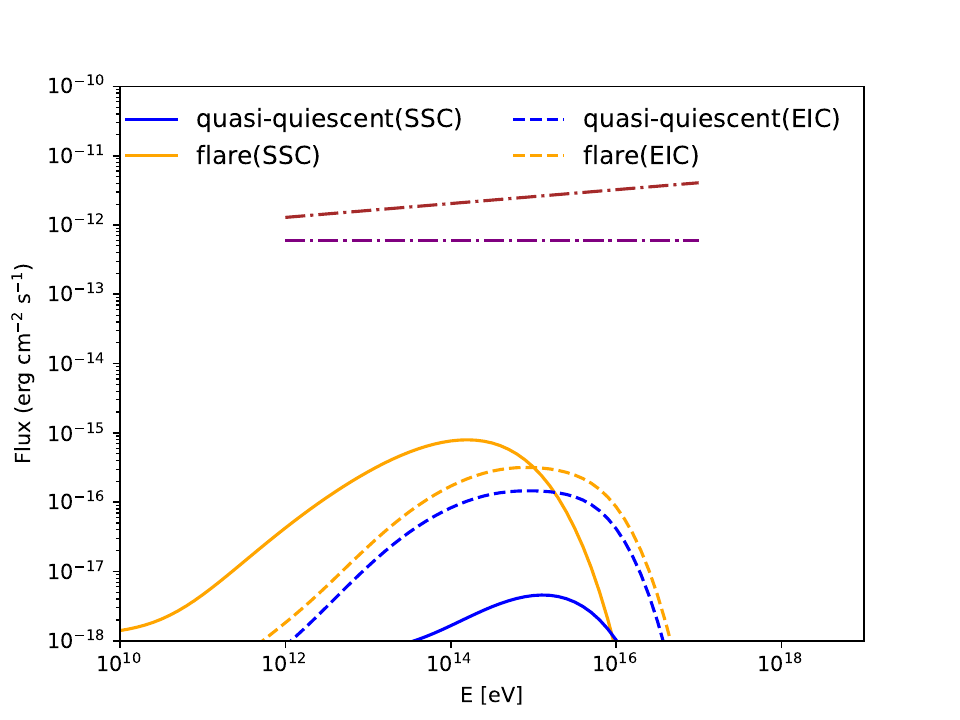}
    \caption{ {Predicted all-flavor neutrino spectra for the SSC and EIC scenarios in the quasi-quiescent  and flaring states. Red and purple dash-dotted lines denote the 90\% upper limit (time-dependent search) and 90\% sensitivity (time-integrated search) from Table~\ref{tab:llagn}.}} 
    \label{fig:leptohadronic}
\end{figure}

The results  {for} the all-flavor neutrino spectrum are summarized in Figure~\ref{fig:leptohadronic}.
Given the IceCube effective area, we calculated the expected number of neutrinos  {for a time window of $T = 58~\rm days$ (flaring state) and $T = 15~\rm yr$ (quasi-quiescent state)}. The number of muon neutrinos (and antineutrinos) can be estimated as 
\begin{equation}
\mathcal{N}_{\nu_\mu} = T\times\int_{E_{\rm th}}^\infty dE~ \mathcal{\phi}_\nu(E) \times A_{\rm eff}(E, {\rm Dec.})  ,
\end{equation}
where $A_{\rm eff}$ is the IceCube effective area at Dec.~= $29.281^\circ$ \citep{2018}, $E_{\rm th}=50\rm~TeV$ is the energy threshold, $\mathcal{\phi}_\nu$ is the modeled neutrino flux.  {The expected number of detections is $ \mathcal{N}_{\nu_\mu} = 5.8\times10^{-4}$ (EIC, quasi-quiescent ), $\mathcal{N}_{\nu_\mu} = 4.0\times10^{-5}$ (SSC, flare), $\mathcal{N}_{\nu_\mu} = 1.4\times10^{-5}$ (SSC, quasi-quiescent),  $\mathcal{N}_{\nu_\mu} = 1.3\times10^{-5}$ (EIC, flare)}.
Thus, it is  {challenging} for IceCube to detect high-energy neutrinos from NGC 4278 based on our model  {in our fiducial scenarios, but the prospect is better in the EIC model in the quasi-quiescent state. Note that the neutrino flux can be higher in the RIAF model, but such efficient neutrino production leads to the requirement that the source should be gamma-ray ``hidden", in which VHE gamma rays observed by LHAASO cannot escape~\citep{Murase:2015xka}.} 

\section{Discussion and summary}\label{sec:sum}
In this study, we conducted a long-term multiwavelength analysis of the radio galaxy NGC 4278 during the 2024–2025 period, identifying the jet in a quasi-quiescent state. We performed numerical modeling of this state and compared it with the 2021 flaring state. Using a one-zone leptonic SSC model combined with MCMC analysis, we derived best-fit parameters and $1\sigma$ uncertainty bands for both epochs. While the origin of the VHE gamma-ray emission in the flaring state is consistent with the SSC model, the model fails to explain the VHE gamma rays in the quasi-quiescent state. Specifically, the SSC model requires a jet power that significantly overshoots observed values, unless a Doppler factor higher than that measured from radio observations is adopted.  {Furthermore, in the SSC model, the electron luminosity during the flaring state is lower than in the quasi-quiescent state, which does not seem natural.}

A  {possible} explanation for the discrepancy between the model predictions and the observations  {for the SSC model} is the use of non-simultaneous multiwavelength data in constructing the SED for the quasi-quiescent state. 
For instance, the average flux of VHE gamma raysobserved in 2024--2025 corresponds to a relatively low  {flux} state.
 {However, the uncertainty introduced by fitting non-simultaneous multiwavelength data is likely to be secondary given the relative stability of the quasi-quiescent state.}
It is also possible that the VHE gamma rays detected by LHAASO during the quasi-quiescent state may receive contributions from other emission components. 
For example, the VHE gamma rays from NGC~4278 could originate from hadronic interactions between cosmic-ray protons and ions that have escaped from the jet and subsequently interacted with surrounding giant molecular clouds~\citep{Shoji:2025znc, Xu:2025ujs}. However, reproducing the LHAASO-observed flux in this scenario would require an enhanced accretion rate by a factor of $\sim 10$--$1000$, and a diffusion coefficient that is $0.1$--$1$ times lower than that inferred for the general interstellar medium, and explaining the variability might be challenging.

 {We showed that EIC emission from the jet provides a viable, more comfortable explanation.} In LL AGNs, variable X-ray emission may be associated with hot accretion flows~\citep{Kimura:2019yjo,Kimura_2021_b,Das:2026}. While it is unlikely that VHE gamma rays escape from the RIAF disk~\citep{Das:2026}, the infrared component from RIAFs can also serve as a promising target photon field for the EIC process in jets or outflows, which successfully explains the observed VHE gamma rays using physically reasonable values for both the Doppler factor and jet power. Our results suggest that the EIC process may play a critical role in the production of VHE gamma rays in LL AGNs.

Finally, we explored a leptohadronic model, comparing the predicted neutrino flux to the observational upper limits derived in this work.
Our results  {show} that the detection of high-energy neutrinos from NGC 4278 is possible assuming roughly 0.1\% of the Eddington luminosity transferred to high-energy protons.
Future multimessenger observations of LL AGNs will be essential to further constrain the physical processes responsible for the production of high-energy gamma rays and neutrinos. 

\begin{acknowledgments}
 {We thank the reviewer for the careful reading of our manuscript and for the valuable and constructive comments.}
We gratefully acknowledge the valuable discussions and insightful comments provided by Shiqi Yu, Feng Qi and Alberto Dominguez.
B.T.Z. is supported in China by National Key R\&D program of China under the grant 2024YFA1611402.
The work of K.M. was supported by the NSF Grant No.~2308021.
S.S.K. acknowledges support by KAKENHI Nos. 22K14028, 21H04487, 23H04899, and  the Tohoku Initiative for Fostering Global Researchers for Interdisciplinary Sciences (TI-FRIS) of MEXT's Strategic Professional Development Program for Young Researchers.
\end{acknowledgments}

\clearpage
\appendix
\section{Details of data analysis result}
We analyzed the Swift-XRT data for NGC 4278.  {The corresponding results are presented in Table \ref{tab:xray_fit_error} and Table\ref{tab:xrt}}. The table~\ref{tab:xrt} lists all the Swift-XRT observations used in this work. The table includes the observation date, observation ID (ObsID), exposure time, and the best-fit model parameters. Additionally, we combined the observational data from 2024 to 2025 to perform the SED fitting. These combined results are also presented in the table.\ref{tab:xray_fit_error}. In fitting model, for the Galactic hydrogen column density, we fixed it to be $2.2 \times10^{20}\rm~cm^{-2}$~\citep{2013MNRAS.431..394W}, and we performed a fit to the excess hydrogen column density, photon index, and flux with C-statistic by \texttt{XSPEC}. We performed an analysis of the multiwavelength light curve~\ref{fig:light_curve}. The results show a simultaneous high state in 2021, with elevated flux in both the X-ray and LHAASO energy range. 
\begin{table}[htbp]
\caption{Best-fit parameters for the Swift-XRT spectrum (2024--2025). Uncertainties correspond to the 90\% confidence level. }
\centering
\begin{tabular}{lcc}
\hline
Parameter & Value & Unit \\
\hline
Galactic $N_\mathrm{H}$ & $0.022$ (fixed) & $10^{22}\ \mathrm{cm}^{-2}$ \\
Excess $N_\mathrm{H}$   & $0.058^{+0.039}_{-0.035}$ & $10^{22}\ \mathrm{cm}^{-2}$ \\
Redshift ($z$)          & 0.00216 & -- \\
Photon Index ($\Gamma$) & $2.11^{+0.17}_{-0.16}$ & -- \\
Obs Flux ($F_\mathrm{obs}$)   & $1.17^{+0.11}_{-0.10}$ & $10^{-12}\ \mathrm{erg\ cm^{-2}\ s^{-1}}$ \\
Unabs Flux ($F_\mathrm{unabs}$) & $1.43^{+0.14}_{-0.11}$ & $10^{-12}\ \mathrm{erg\ cm^{-2}\ s^{-1}}$ \\
C-stat / DOF            & $252.9 / 283$ & -- \\
Goodness ($\chi^2$)     & 268.0 & -- \\
Net Count Rate          & $0.027$ & $\mathrm{counts\ s^{-1}}$ \\
Exposure Time           & 26.8 & ksec \\
\hline
\end{tabular}
\label{tab:xray_fit_error}
\end{table}

\begin{table*}[htbp]
\caption{\label{tab:xrt} Log of Swift-XRT observations of NGC 4278 and spectral fitting results. 
Fluxes are in the 0.3--10.0 keV band. The Galactic column density was fixed at $N_{\rm H} = 2.22 \times 10^{20}\,\mathrm{cm^{-2}}$.
Errors correspond to the 90\% confidence level.
}
\begin{ruledtabular}
\begin{tabular}{lccccc}
        Date(UT) & ObsID & Exposure & $\Gamma$ &$N_{H}({\rm Intrinsic})$ &Flux(0.3-10.0 keV) \\ 
       & &(s) & & $(10^{20} \rm{cm^{-2}}$) & ($10^{-12}\rm{erg/cm^{2}/s}$)
        \\
        \hline
          2021-02-26 11:13:35&03109562001   &173.0  &--      &--    &-- \\ 
          2021-11-28 20:16:34&03109562002   &922.7  &$1.2^{+0.7}_{-0.3}$&$0^{+17.9}_{-0}$&$4.5^{+1.6}_{-1.7}$\\ 
          2024-05-31 07:47:56&00016648001   &1662.3 &$2.3^{+0.7}_{-0.4}$&$0^{+0.12}_{-0}$ &$0.89^{+0.3}_{-0.25}$ \\ 
          2024-06-03 05:21:54&00016648002   &1632.3  &--     &--    &-- \\ 
          2024-06-06 09:03:56&00016648003   &1559.5  &$1.9^{+0.7}_{-0.5}$   &$6.2^{+19.4}_{-6.2}$ &$1.4^{+0.7}_{-0.4}$ \\ 
          2024-12-13 03:54:29&00016648004   &3638.1  &$2.0^{+0.6}_{-0.3}$&$0.19^{+11.82}_{-0.19}$&$0.78^{+0.19}_{-0.21}$ \\ 
          2024-12-16 03:56:57&00016648005   &2033.4  &$1.9^{+0.6}_{-0.5}$&$11^{+20}_{-11}$&$1.4^{+0.6}_{-0.4}$ \\ 
          2024-12-19 01:02:56&00016648006   &4360.2  &$2.2^{+0.4}_{-0.3}$&$3^{+9}_{-3}$ &$1.08^{+0.25}_{-0.21}$ \\ 
          2024-12-22 01:10:57&00016648007   &4327.6  &$1.79^{+0.36}_{-0.39}$&$3.8^{+10.1}_{-3.8}$& $1.8^{+0.4}_{-0.3}$ \\ 
          2024-12-25 03:10:57&00016648008   &4415.4  &$2.1^{+0.4}_{-0.3}$&$4^{+9}_{-4}$&$1.33^{+0.32}_{-0.25}$ \\ 
          2024-12-06 22:50:25&00089850001   &1575.8  &--      &--    &-- \\ 
          2025-01-13 17:12:55&00089850002   &1582.1  &$2.4^{+0.9}_{-0.8}$ &$0.17^{+0.27}_{-0.17}$&$1.3^{+0.7}_{-0.4}$\\ 
\end{tabular}
\end{ruledtabular}
\end{table*}

\section{Fitting results of SSC model\label{coutour-fig}}
\begin{figure}[h]
\includegraphics[width=0.45\linewidth]{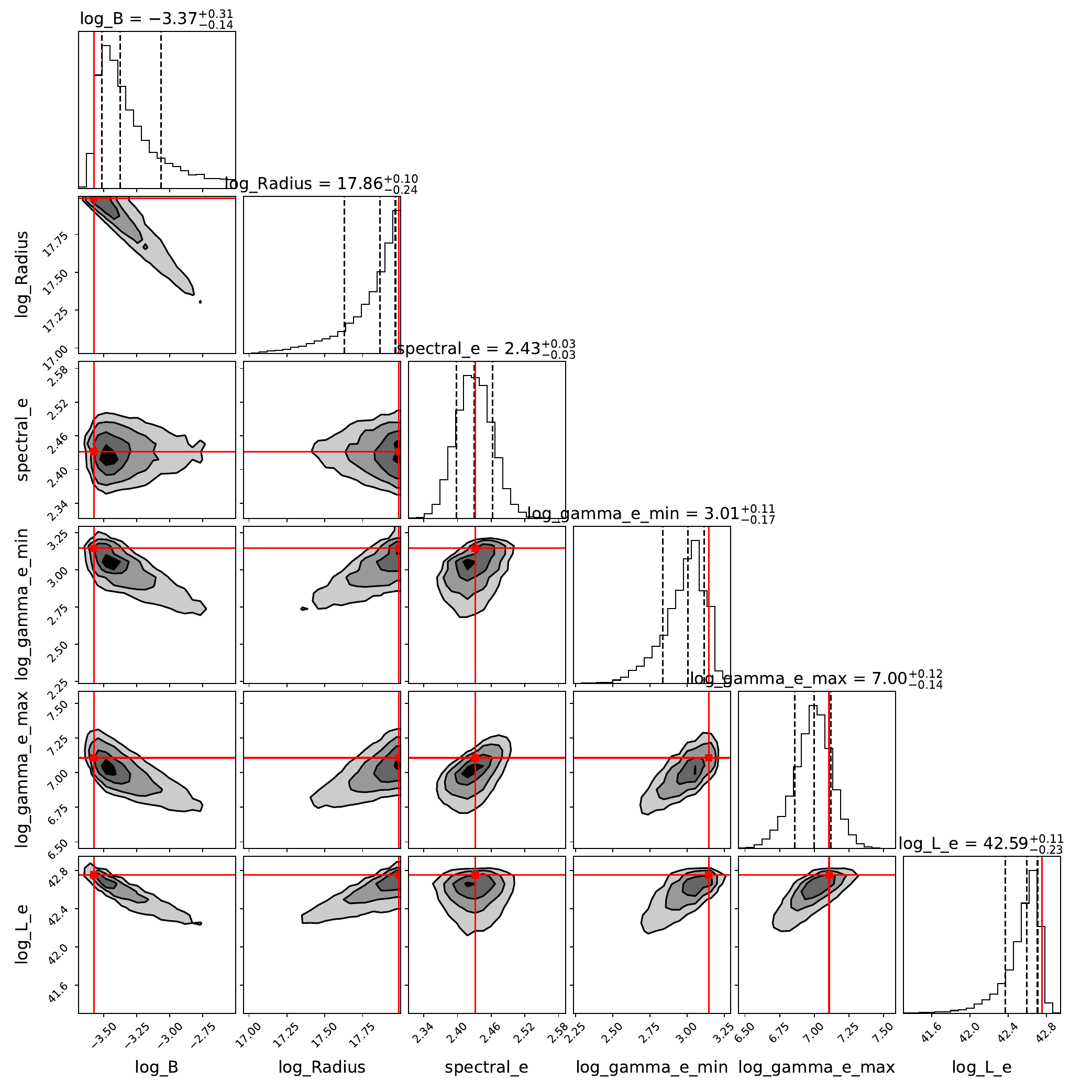}
\includegraphics[width=0.45\linewidth]{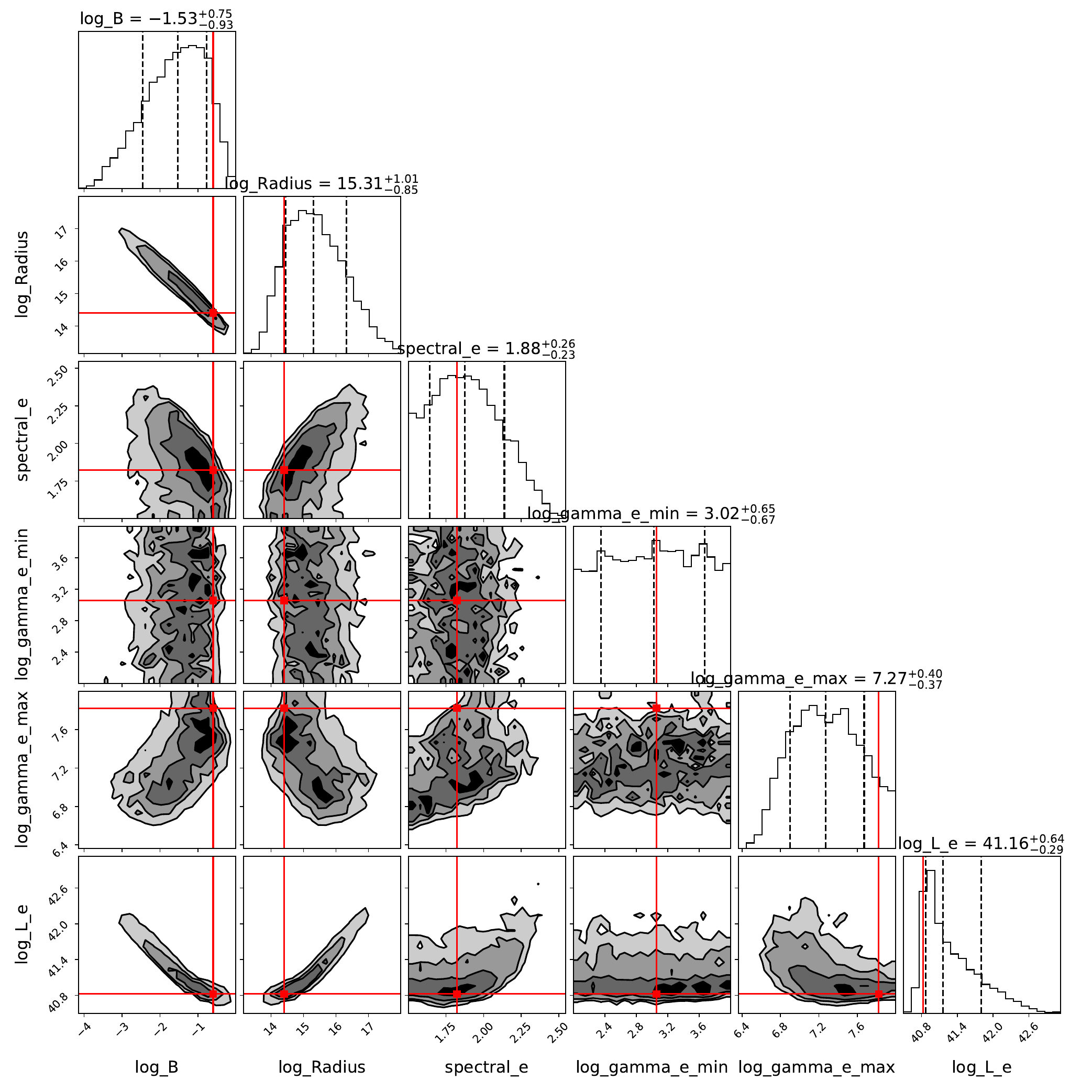}

\caption{Contour plot in the quasi-quiescent  state (left panel) and flaring state (right panel) for fixed Doppler factor $\delta$, respectively.}
\label{fig:contour-fixed}
\end{figure}

\begin{figure}[htbp]
\includegraphics[width=0.45\linewidth]{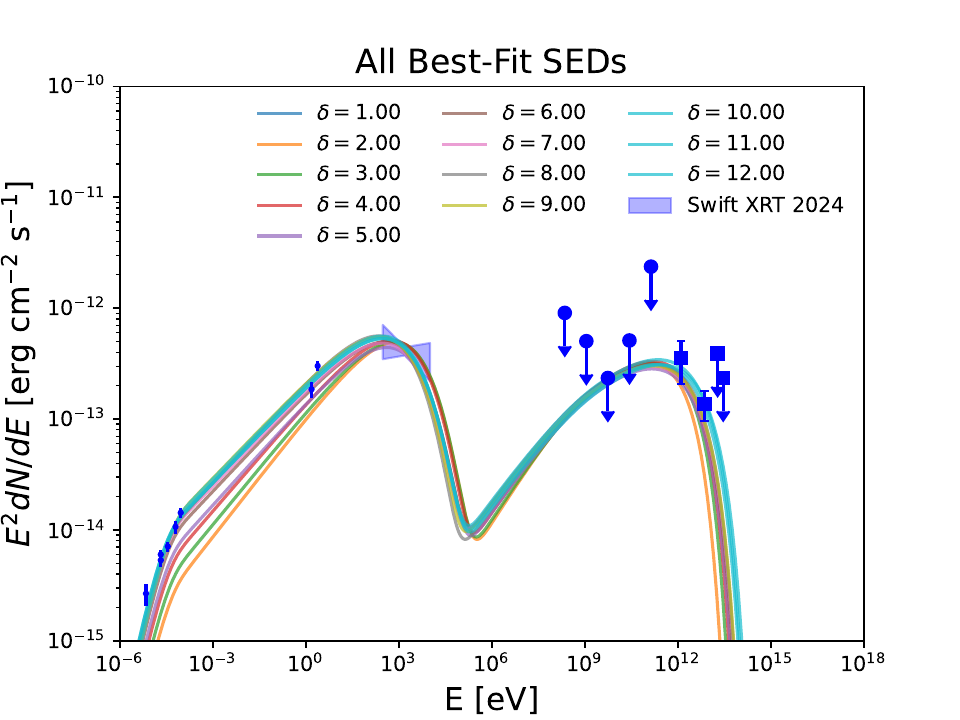}
\includegraphics[width=0.45\linewidth]{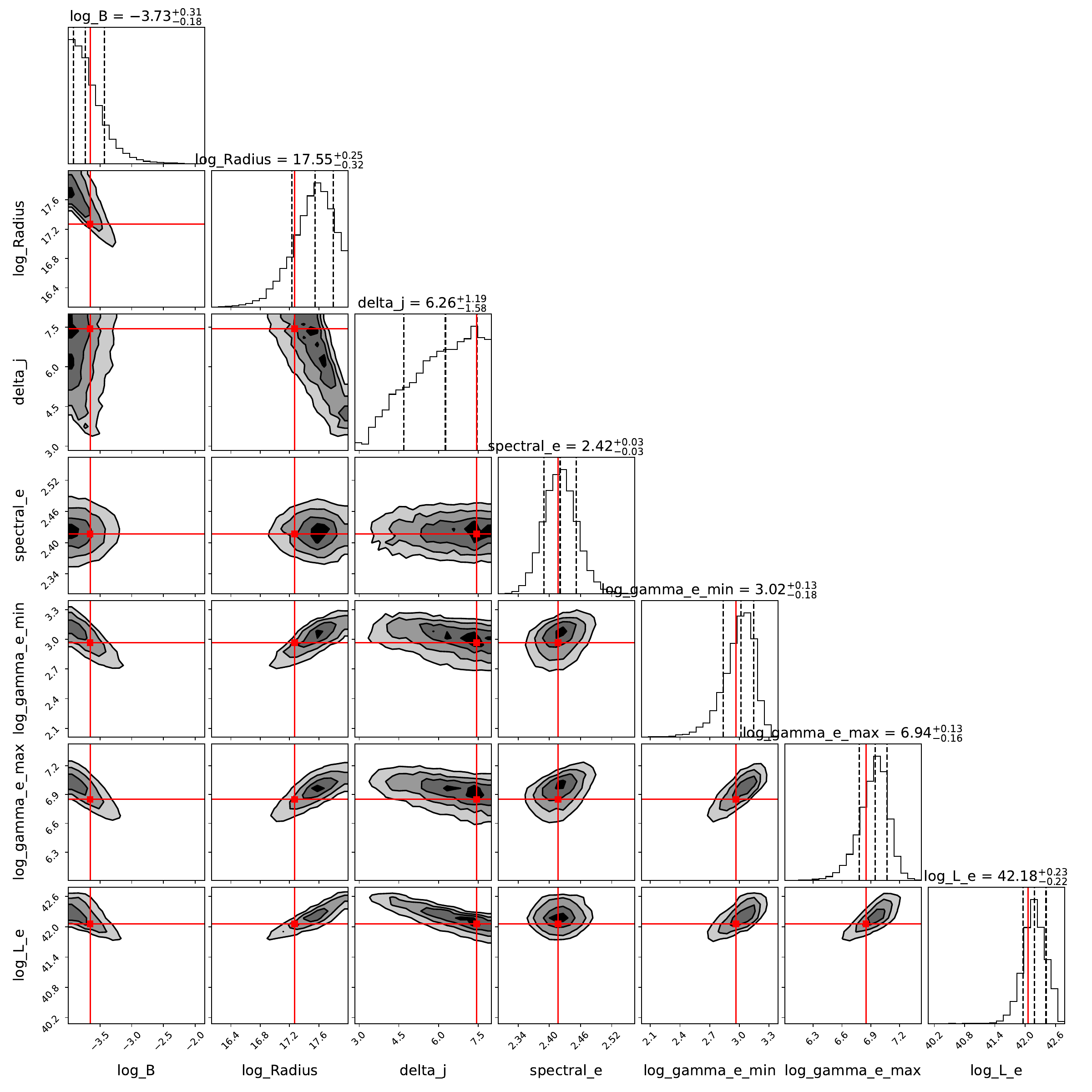}
\caption{Left: The marginalized maximum-likelihood method with the variation of Doppler factor $\delta$. Right: Contour plot in the quasi-quiescent  state with $\delta$ as free parameter.}
\label{fig:contour-free}
\end{figure}

Here, we present the contour plots of the fitting results in the one-zone SSC model in Figure~\ref{fig:contour-fixed}. 
In Figure~\ref{fig:contour-free}, we present the fitting results with the Doppler factor as a free parameter in both marginalized maximum-likelihood method (left panel) and MCMC fitting (right panel).

\clearpage
\bibliography{main}{}
\bibliographystyle{aasjournalv7}

\end{document}